\documentclass[a4paper,11pt]{article}
\usepackage{pos}

\usepackage{slashed}

\title{Pion condensation at lower than physical quark masses}

\author[a]{Bastian B. Brandt}
\author*[b]{Volodymyr Chelnokov}
\author[b]{Francesca Cuteri}
\author[a]{Gergely Endr\H{o}di}

\affiliation[a]{
  Institute for Theoretical Physics, University of Bielefeld, \\
  D-33615 Bielefeld, Germany
}
\affiliation[b]{
  Institut f\"{u}r Theoretische Physik, Goethe-Universit\"{a}t Frankfurt\\
 Max-von-Laue-Str.\ 1, 60438 Frankfurt am Main, Germany
}

\emailAdd{brandt@physik.uni-bielefeld.de}
\emailAdd{chelnokov@itp.uni-frankfurt.de}
\emailAdd{cuteri@itp.uni-frankfurt.de}
\emailAdd{endrodi@physik.uni-bielefeld.de}

\abstract{
In QCD at large enough isospin chemical potential Bose-Einstein Condensation (BEC) takes place, separated from the normal phase by a phase transition. From previous studies the location of the BEC line at the physical point is known. In the chiral limit the condensation happens already at infinitesimally small isospin chemical potential for zero temperature according to chiral perturbation theory. The thermal chiral transition at zero density might then be affected, depending on the shape of the BEC boundary, by its proximity. As a first step towards the chiral limit, we perform simulations of 2+1 flavors QCD at half the physical quark masses. The position of the BEC transition is then extracted and compared with the results at physical masses.
}

\FullConference{%
  The 39th International Symposium on Lattice Field Theory (Lattice2022),\\
  8-13 August, 2022 \\
  Bonn, Germany 
}

\DeclareMathOperator{\Tr}{\rm Tr}

\begin{document}
\maketitle

\section{Introduction}

The phase structure of QCD continues to be an object of 
great interest for both experimental and theoretical studies.
There is currently a wealth of numerical simulation results 
at zero quark density, 
from which we know, in particular, that the chiral symmetry restoration at physical quark masses is a continuous crossover \cite{physical-crossover}. 
Studying the theory in the chiral limit is complicated
by the appearance of zero modes of the Dirac operator, 
and the need to go to the continuum limit to restore chiral symmetry 
with Wilson or staggered fermions, 
or to employ computationally much costlier chiral symmetric fermion discretizations. 
Nevertheless, there is a recent progress in this region which, in particular, suggests that in the limit $m_{\rm ud} = 0$ the chiral transition is of second order (for $N_f = 2,3$) and 
becomes a crossover at arbitrarily small nonzero quark masses~\cite{chiral-second-order}.

Extending the QCD phase diagram to nonzero densities is confounded by 
the sign problem -- the Boltzmann weight of the lattice configuration becomes complex at nonzero $\mu_B$, preventing straightforward use of importance sampling. Different ways to overcome this problem 
were proposed (see \cite{sign-problem-review-Forcrand, sign-problem-review-Nagata} for a review), but still the problem is far from being solved.

\begin {figure}[b]
\centering
\includegraphics[width=0.8\textwidth,clip]{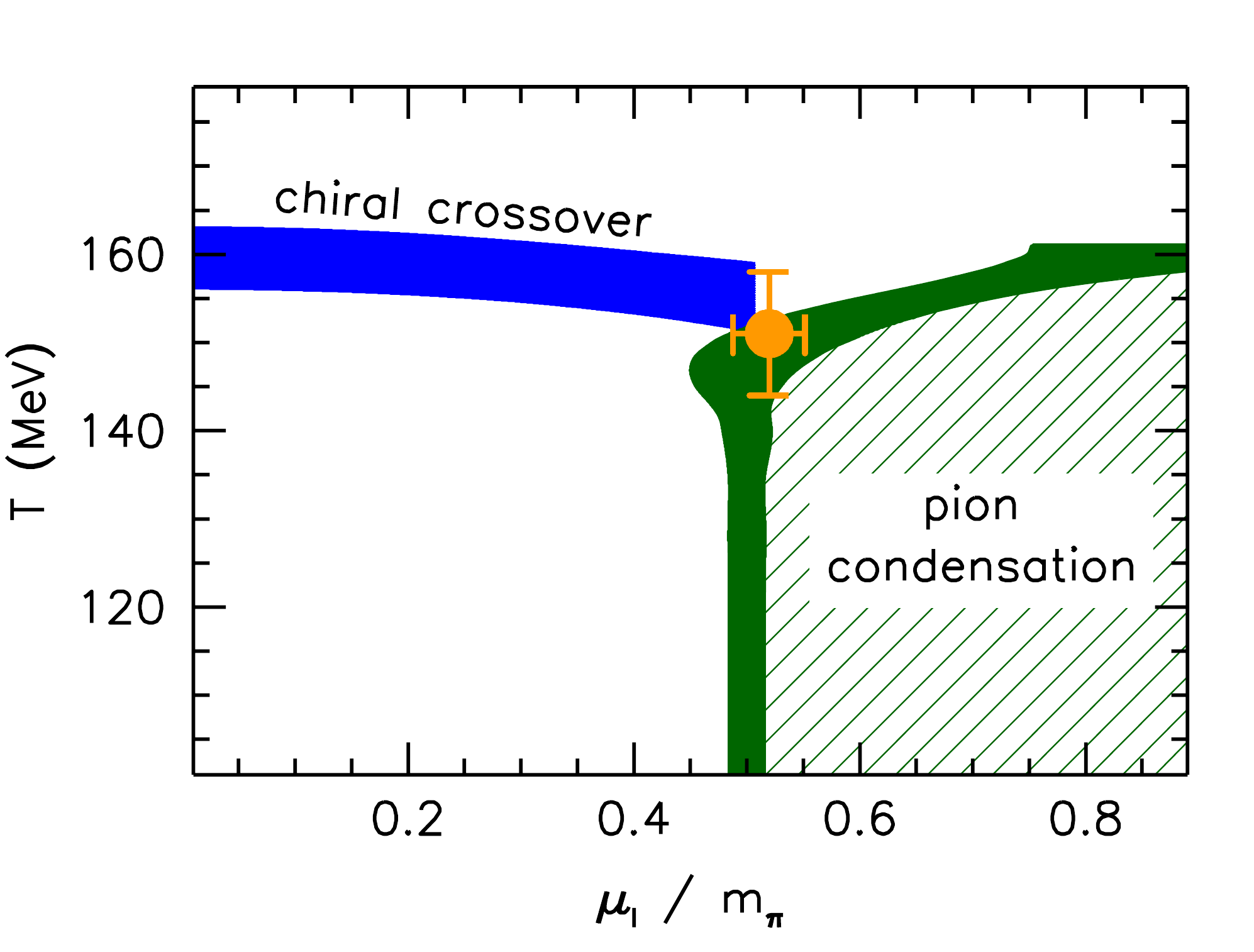}
\caption{The QCD phase diagram in $T$-$\mu_I$ plane at physical quark masses, obtained in~\cite{muI-phys}.
}
\label{fig:PD_phys}
\end{figure}

Another possible extension of the QCD phase diagram is to 
the region of nonzero isospin density at $\mu_I \neq 0$, potentially relevant for early Universe cosmology~\cite{Vovchenko:2020crk}, for instance.
At pure isospin chemical potential $\mu_I \neq 0, \mu_B = 0$, 
the theory is sign problem-free and the standard Monte-Carlo simulations can be performed~\cite{finite-isospin}. The recent study \cite{muI-phys} 
shows that the deconfinement transition remains a crossover for 
$\mu_I \le m_\pi / 2$,
until it intersects with the second order pion condensation line, which is found to be approximately vertical at small enough $T$ (see Figure~\ref{fig:PD_phys}). The $\mu_I$-$T$ plane can also be used to check the methods targeted at 
studying QCD at finite $\mu_B$ -- since we can compare their results with the results of direct simulation, that is possible here \cite{taylor-reliability}. Additionally, the ability to numerically sample the theory in the 
$\mu_I$-$T$ plane can be exploited to perform the reweighting to nonzero $\mu_B$ \cite{isospin-reweighting}.

\begin{figure}[ht]
\centering
\includegraphics[scale=0.46]{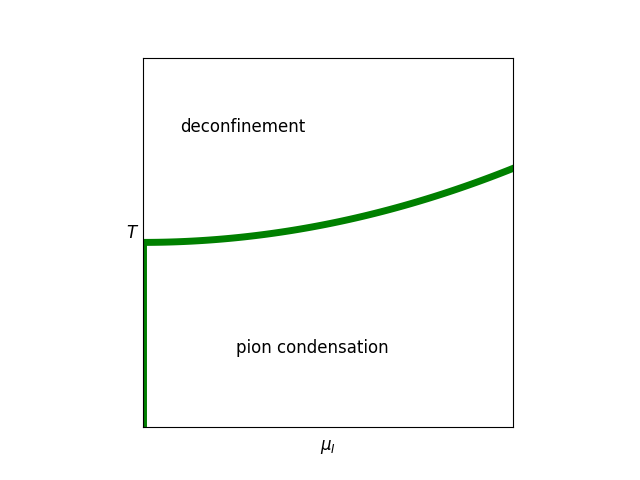} \includegraphics[scale=0.46]{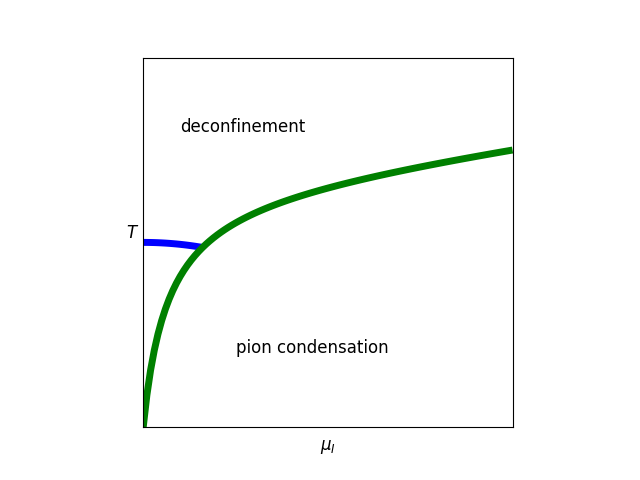} 
\\
\caption{Two possibilities for the phase diagram in $T$-$\mu_I$ plane 
at the chiral limit.} 
\label{fig:chiral-pictures}
\end{figure}

The phase diagram shown in Figure~\ref{fig:PD_phys} has an interesting implication concerning the chiral phase diagram. At zero temperature, the
pion condensation happens at $\mu_I = m_\pi / 2$. When the light quark mass goes to zero, the pion mass also goes to zero as $m_\pi\sim\sqrt{m_{\rm ud}}$ according to chiral perturbation theory. 
Thus at least at zero temperature in the chiral limit pion condensation happens already at arbitrarily small isospin chemical potential. 
From Figure~\ref{fig:PD_phys} we see that at physical mass the 
isospin transition line remains vertical up until it meets the chiral crossover line. If that holds also at $m_{\rm ud} = 0$, then in the chiral limit the pion condensation line would lie on the $\mu_I = 0$ axis up to $T = T_c$ -- i.e.\ up to the chiral phase transition temperature. In this scenario the pion condensate in the chiral limit at $T=T_c$ exists at arbitrarily small isospin chemical potential, affecting the chiral phase transition at zero chemical potential (Figure~\ref{fig:chiral-pictures}, left). 
Alternatively, the phase transition line can start bending towards larger $\mu_I$ as the quark masses are reduced, 
resulting in a different phase structure (Figure~\ref{fig:chiral-pictures}, right). 
Previous studies using the Nambu-Jona-Lasinio model \cite{NJL-chiral-isospin}, as well as the functional renormalization group \cite{FRG-chiral-isospin}, support the first phase picture. To verify this, 
direct Monte Carlo QCD simulations are necessary.

As a first step, in this work we perform the simulation of QCD 
at nonzero isospin density for $m_{\rm ud} = m_{\rm ud, phys} / 2$, comparing the position of the pion condensation transition with 
the results at physical masses at a series of temperatures.

\section{Simulation setup} 

We study $2+1$-flavour QCD using 
staggered fermions on a $24^3 \times 8$ lattice. 
The setup is the same as the one used in~\cite{muI-phys}, with the light quark masses changed to half their physical value. 
The partition function of the theory has the form
 \begin{equation}
\mathcal{Z} = \int \mathcal{D} U_\mu \, 
    e^{-\beta S_G}\, 
    (\det \mathcal{M}_{\rm ud})^{1/4}\,
    (\det \mathcal{M}_{\rm s})^{1/4} \,,
\label{partition-function}
\end{equation}
where $S_G$ is the tree-level Symanzik improved gauge action, $\mathcal{M}_{\rm ud}$ and $\mathcal{M}_{\rm s}$ are, correspondingly, light and strange quark matrices
\begin{equation}
    \mathcal{M}_{\rm ud} = 
\begin{pmatrix}
  \slashed{D}(\mu_I) + m_{\rm ud} & \lambda \eta_5 \\
 -\lambda \eta_5 & \slashed{D}(-\mu_I) + m_{\rm ud}
\end{pmatrix}\,, \;\;\,
\mathcal{M}_s = \slashed{D}(0) + m_s\,.
\label{quark-matrices}
\end{equation} 
Here $\lambda$ is a pion source term which explicitly breaks
the $U_{\tau_3}(1)$ symmetry of the action, 
which both allows us to see the pion condensate 
on the finite volume lattice, and, at the same time, 
improves the condition number of the light quark matrix. 
To get physically meaningful results, the limit $\lambda \to 0$ 
must be taken.

The Boltzmann weight defined by (\ref{partition-function}) is positive, 
which is obvious for the gauge action and the strange quark determinant. 
The positiveness of the light quark determinant follows from the 
generalized $\eta_5$-hermiticity relation that the Dirac operator satisfies,
\begin{equation}
 \eta_5 \slashed{D}(-\mu_I)\eta_5 = \slashed{D}(\mu_I)^\dagger \ .
 \label{generalized-hermiticity}
\end{equation}
Now, taking the matrix $B=\mathrm{diag}(1, \eta_5)$ in flavour space, 
that has a unit determinant, we get
\begin{align}
B \mathcal{M}_{\rm ud} B &= \begin{pmatrix}
  \slashed{D}(\mu_I) + m_{\rm ud} & \lambda \\
 -\lambda & \left(\slashed{D}(\mu_I) + m_{\rm ud}\right)^\dagger
\end{pmatrix} \ , \nonumber \\
\det \mathcal{M}_{\rm ud} &= \det \left(B \mathcal{M}_{\rm ud} B\right) = 
\det \left( \left|\slashed{D}(\mu_I) + m_{\rm ud}\right|^2 + \lambda^2 \right) > 0 \ .
\label{determinant-positivity}
\end{align}

To locate the pion condensation onset, we measure the pion condensate 
\begin{equation}
 \Sigma_{\pi} = \frac{m_{ud}}{m_\pi^2 f_\pi^2}  \left\langle \pi^\pm \right\rangle \ ,
 \label{renormalized-pion-condensate}
\end{equation}

\begin{equation}
  \left\langle \pi^\pm \right\rangle 
= \frac{T}{V}\frac{\partial \log \mathcal{Z}}{\partial \lambda} 
= \frac{T}{2V} \left\langle \Tr \frac{\lambda}{|\slashed{D}(\mu_I)+m_{\rm ud}|^2+\lambda^2} \right\rangle \ ,
 \label{pion-condensate}
\end{equation}
where the trace can be calculated using noisy estimators. 
The multiplicative renormalization term in (\ref{renormalized-pion-condensate}) depends on the light quark mass $m_{ud}$, the pion mass $m_{\pi}$ that corresponds to $m_{ud}$, and the pion decay constant $f_{\pi}$.
The chiral perturbation theory predicts that $m_\pi \sim \sqrt{m_{ud}}$
when the light quark mass goes to zero, so assuming this relation holds
for physical light quark mass we can approximate the renormalization (\ref{renormalized-pion-condensate}) as
\begin{equation}
 \Sigma_{\pi} = \frac{m_{ud, {\rm phys}}}{m_{\pi, {\rm phys}}^2 f_\pi^2}  \left\langle \pi^\pm \right\rangle \ ,
 \label{renormalized-pion-condensate-physical}
\end{equation}
where $m_{ud, {\rm phys}}$ and $m_{\pi, {\rm phys}}$ are physical light quark 
and pion masses.

\begin{figure}[ht]
\centering
\includegraphics[width=0.7\textwidth,clip]{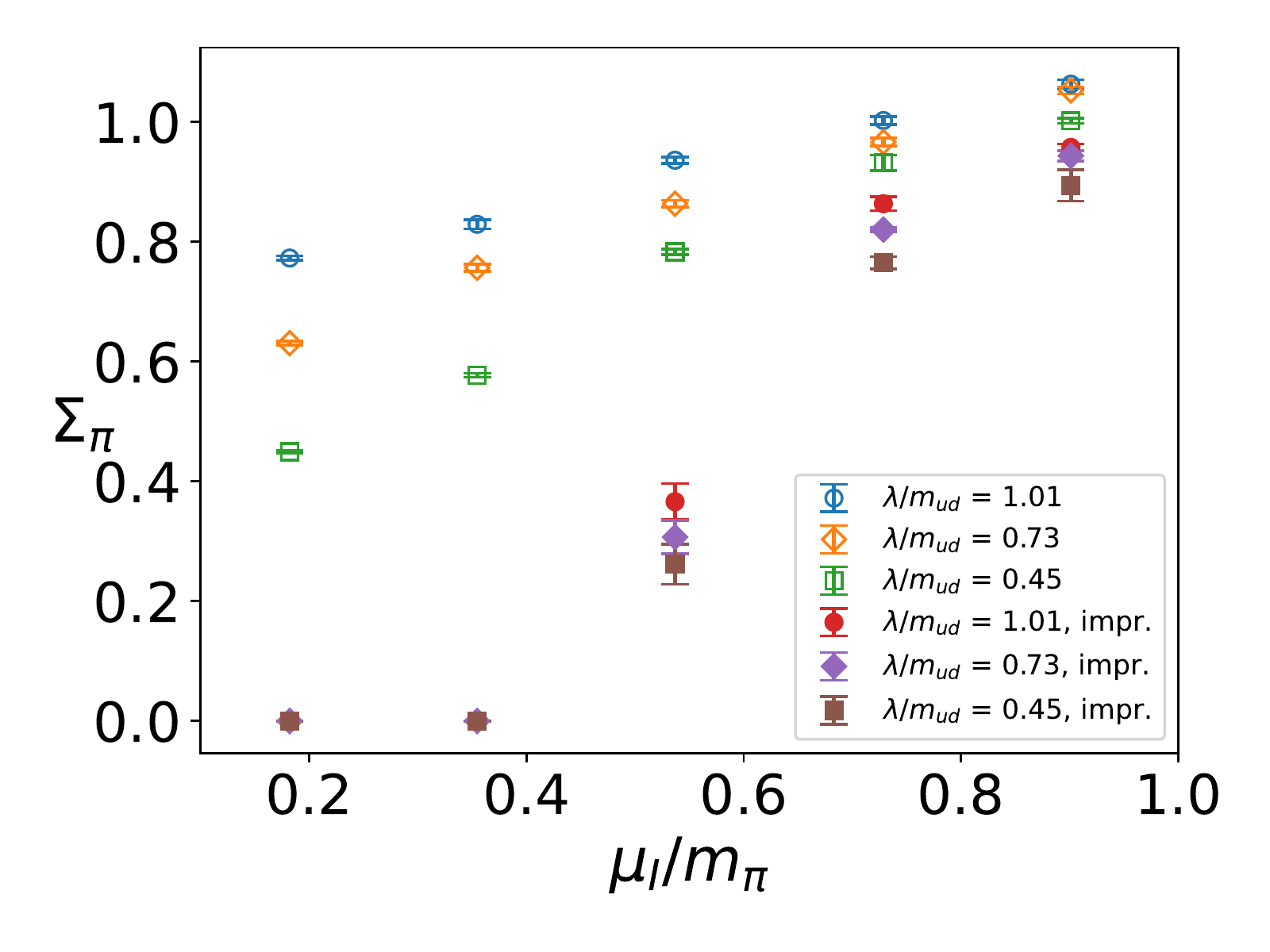}
\caption{Comparison of the improved and the unimproved pion condensate at $T=114$ MeV for three different values of $\lambda$.
}
\label{fig:unimproved-condensate}
\end{figure}

The pion condensate estimated using Eq.~\eqref{renormalized-pion-condensate} 
at $T=114$ MeV, for three different values of $\lambda$ can be seen in
Figure~\ref{fig:unimproved-condensate}. We see that even for the smallest value of $\lambda$ we are still far from the limiting regime $\lambda \to 0$. Unfortunately, the simulations at smaller $\lambda$ values become prohibitively expensive due to very large condition numbers of the 
light quark matrix, that causes a sharp increase of iterations needed
for matrix inversion. At even smaller $\lambda$ values this results in 
a loss of convergence of the conjugate gradient method. 

\section{Improved pion condensate}

We can improve the convergence to the $\lambda \to 0$ limit by using
the Banks-Casher type relation for the pion condensate (\ref{pion-condensate}), that was obtained in~\cite{muI-phys}
\begin{equation}
\left\langle \pi^\pm \right\rangle = 
\frac{T}{2V} \left\langle \Tr \frac{\lambda}{|\slashed{D}(\mu_I)+m_{\rm ud}|^2+\lambda^2} \right\rangle  
= \frac{T}{2V} \left\langle \sum_n \frac{\lambda}{\xi_n^2+\lambda^2} \right\rangle\ ,
\label{singular-sum}
\end{equation}
where $\xi_n$ is the $n$-th singular value of the Dirac operator:
\begin{equation}
\left(\slashed{D}(\mu_I) + m_{\rm ud}\right)^\dagger \left(\slashed{D}(\mu_I) + m_{\rm ud}\right) \phi_n = 
\xi_n^2 \phi_n \ .
\label{singular-def}
\end{equation}
In the infinite volume limit, the summation can be replaced by integration
\begin{equation}
\left\langle \pi^\pm \right\rangle = 
\frac{\lambda}{2} \left\langle \int \textmd{d} \xi \,\rho(\xi) (\xi^2+\lambda^2)^{-1} \right\rangle\ ,
\label{singular-int}
\end{equation}
and taking the limit $\lambda \to 0$ we get
\begin{equation}
\left\langle \pi^\pm \right\rangle = 
\frac{\pi}{4} \left\langle \rho(0) \right\rangle\ .
\label{density-at-zero}
\end{equation}
Thus, the limit $\lambda \to 0$ of the pion condensate is proportional to 
the singular value density at zero. 
Due to the finite size of the lattice the singular values form a discrete 
set, but we can approximate the density by calculating the fraction of the
singular values lying in the interval $[0, \xi]$, and then extrapolating to 
$\xi = 0$ 
\begin{equation}
\rho(0) = \frac{T}{V} \lim_{\xi \to 0} \frac{n(\xi)}{\xi} \ ,
\label{rho-extrapolation}
\end{equation}

\begin{figure}[ht]
\centering
\includegraphics[scale=0.46]{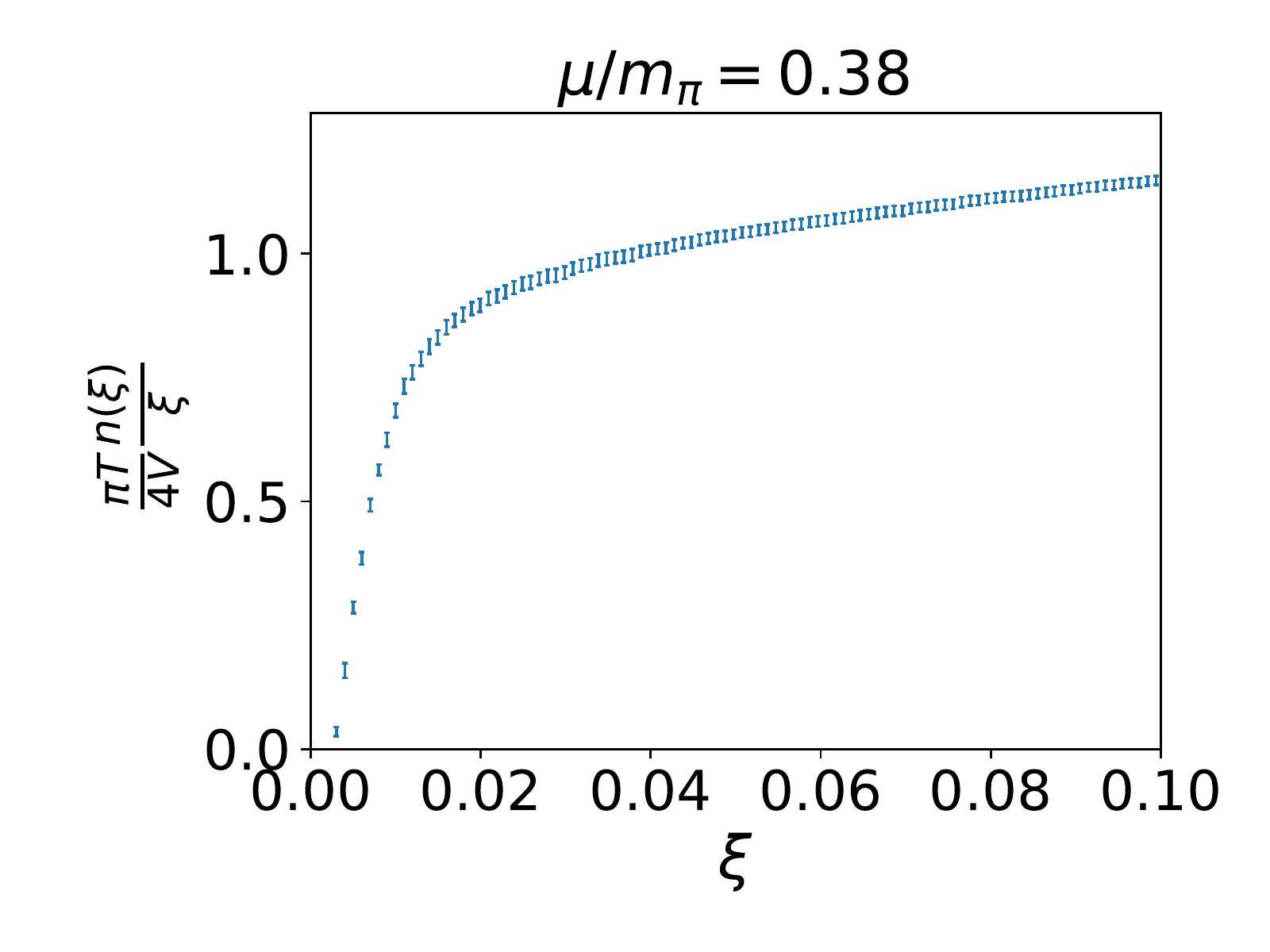} \includegraphics[scale=0.46]{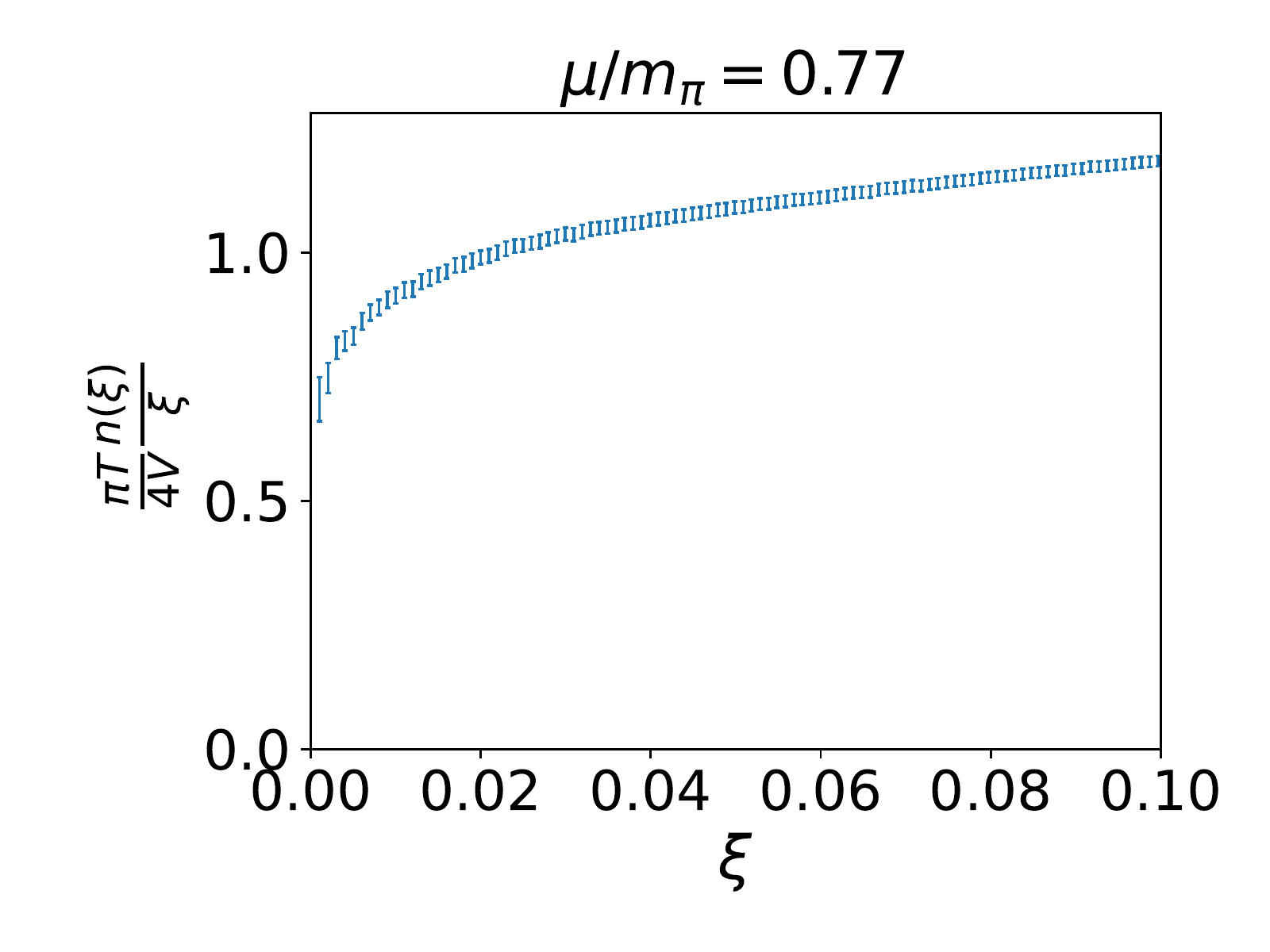} \\
\includegraphics[scale=0.46]{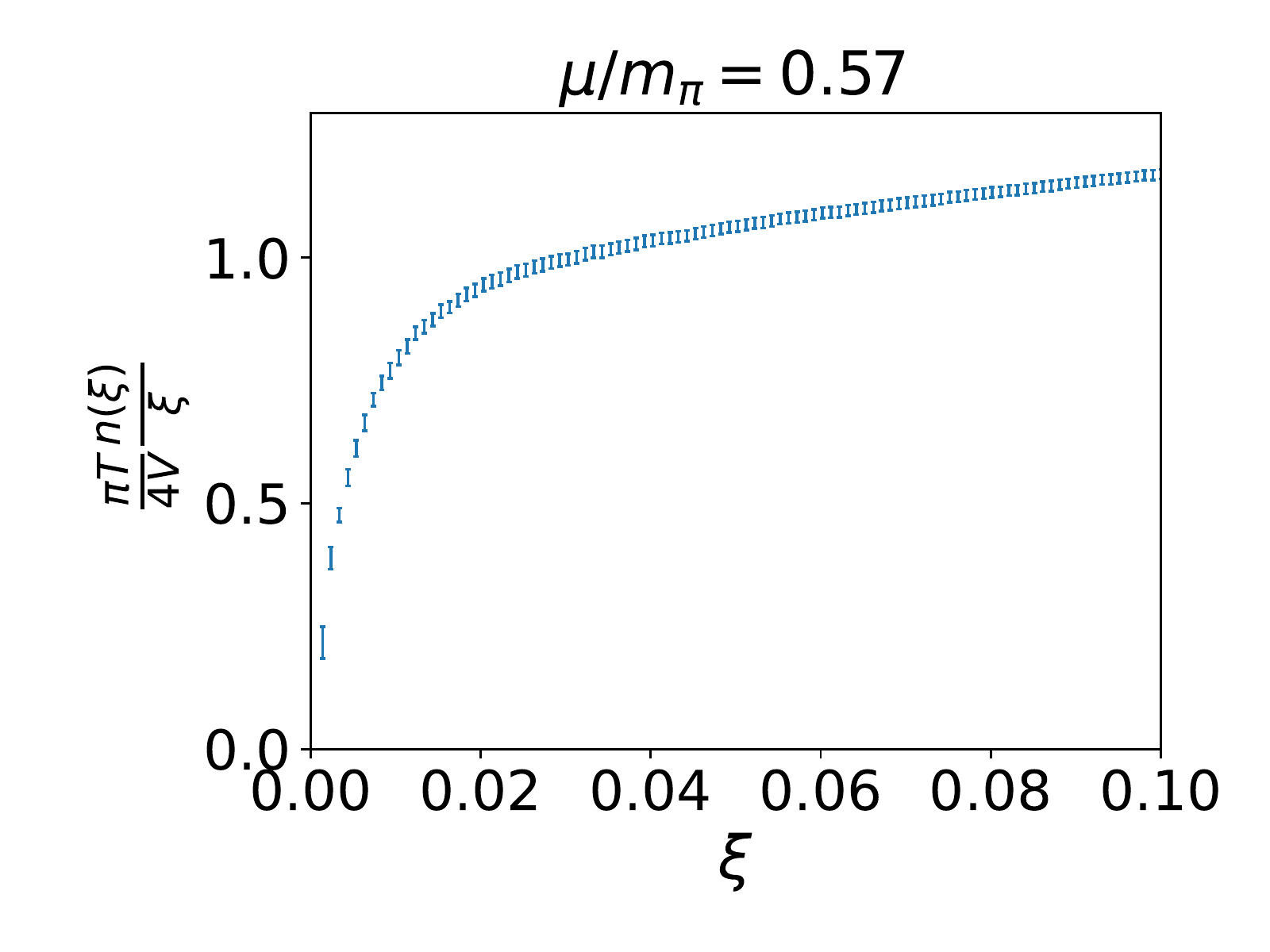}
\caption{The integrated spectral density dependence on the integration region $\xi$ for $\mu_I / m_\pi < 0.5$ (left), $\mu_I / m_\pi > 0.5$ (right), and $\mu_I / m_\pi \approx 0.5$ (bottom) ($T=132$ MeV).}
\label{fig:singular-densities}
\end{figure}

The behavior of such ``averaged densities'' for three regimes -- finite
spectral gap at $\mu_I / m_\pi < 0.5$, $\rho(0) \approx 0$ at $\mu_I / m_\pi \approx 0.5$, and $\rho(0) > 0$ at $\mu_I / m_\pi > 0.5$, is shown on Figure~\ref{fig:singular-densities}.

To perform the extrapolation to $\xi = 0$ we perform a polynomial fit of the integrated spectral density in a given region, obtaining a value of $\left\langle\rho(0)\right\rangle$ and then take a weighted median of all the fit results with the weight $\exp(-\chi^2_r)$, 
where $\chi^2_r$ is the (correlated) chi-squared statistic of the fit per degree of freedom. 
 The error estimate of $\left\langle\rho(0)\right\rangle$ is then taken as a value $\Delta \rho$, for which the interval $[\left\langle\rho(0)\right\rangle - \Delta \rho,
\left\langle\rho(0)\right\rangle + \Delta \rho]$ contains 0.68 of the 
total weight of the estimates.

We can further improve the convergence of the improved pion condensate observable to the limit $\lambda \to 0$ by approximating $ \left\langle \rho(0) \right\rangle_{0}$ (the expectation value of $\rho(0)$ 
with respect to the $\lambda=0$ partition function) using the leading order reweighting, expanding the Boltzmann weight 
of a given configuration in a Taylor series in $\lambda$,
\begin{equation}
 \left\langle \rho(0) \right\rangle_{0} = 
 \frac{\left\langle \rho(0) W(\lambda) \right\rangle_{\lambda}}{\left\langle W(\lambda) \right\rangle_{\lambda}} \ , \quad 
 W(\lambda) = \left( \frac{\det \mathcal{M}_{ud, 0}}{\det \mathcal{M}_{ud, \lambda}} \right)^{1/4} 
  \approx \exp \left( - \frac{\lambda V}{2 T} \pi^\pm \right) \ .
 \label{leading-order-reweighting}
\end{equation}

\begin{figure}[ht]
\centering
\includegraphics[scale=0.7]{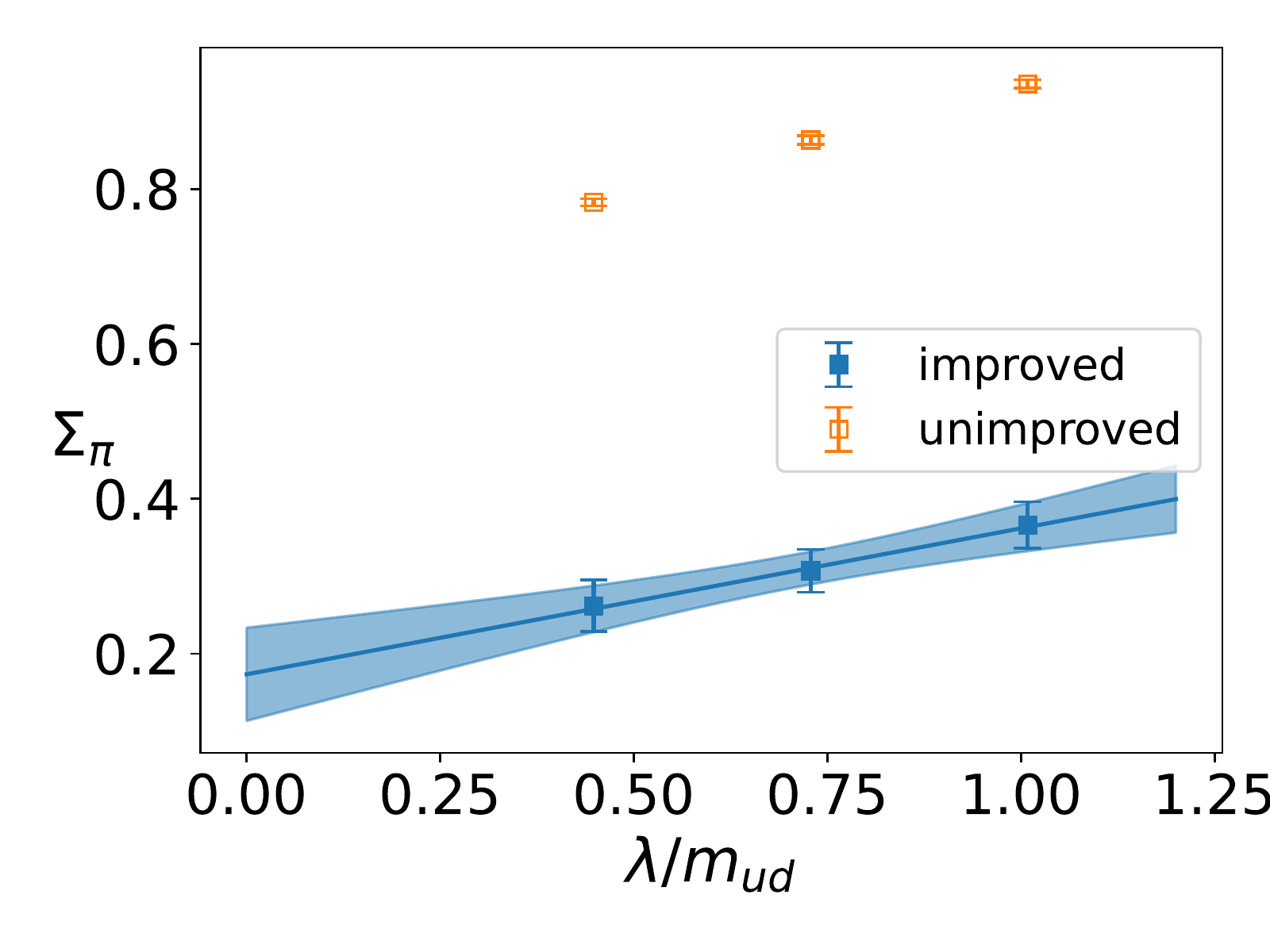} 
\caption{Extrapolation of the improved and unimproved pion condensate to 
$\lambda=0$ at $T=114$ MeV, $\mu/m_\pi = 0.54$. 
}
\label{fig:lambda-extrapolation}
\end{figure}

In Figure~\ref{fig:lambda-extrapolation} we show the 
values of improved and unimproved pion condensate at the point close to $\mu/m_\pi = 0.5$ together with an extrapolation of the improved condensate 
to $\lambda=0$. We can see that the improved pion condensate value is much smaller and less sensitive to the $\lambda$ than the unimproved condensate. 
This picture can be compared with the extrapolaton done in Figure 6 (top) in~\cite{muI-phys}, which has a similar behavior for large $\lambda$, 
but also provides enough data points at small $\lambda$ to actually 
perform the extrapolation of both the unimproved and improved pion condensates
and confirm that they reach the same value at $\lambda=0$. 
Since in that study the linear extrapolation of the improved pion condensate 
worked well for $\lambda/m_{ud} \leq 1.3$, we rely on it also in this study.


\section{Results and summary}

\begin{figure}[ht]
\centering
\includegraphics[scale=0.44]{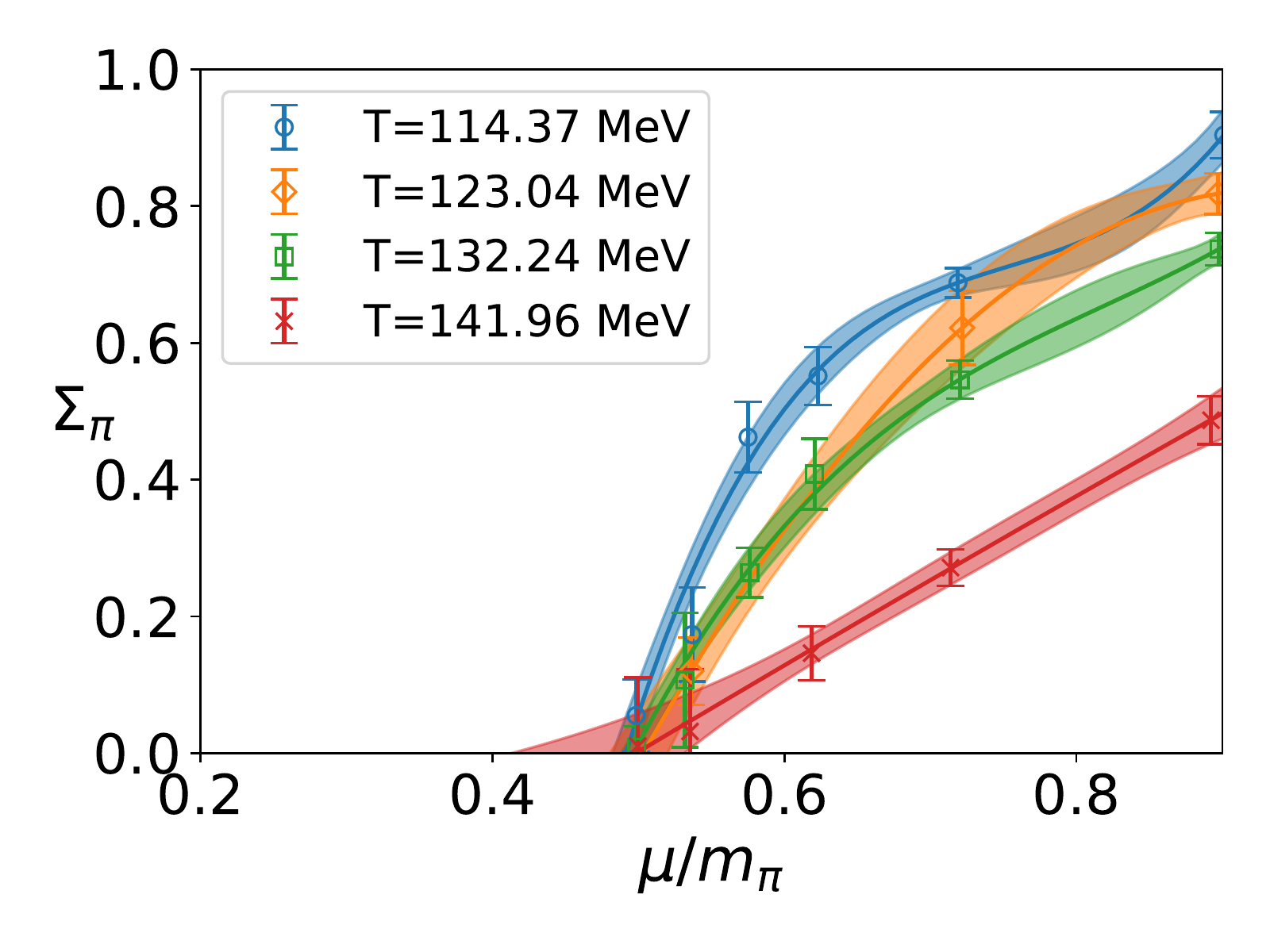} \includegraphics[scale=0.44]{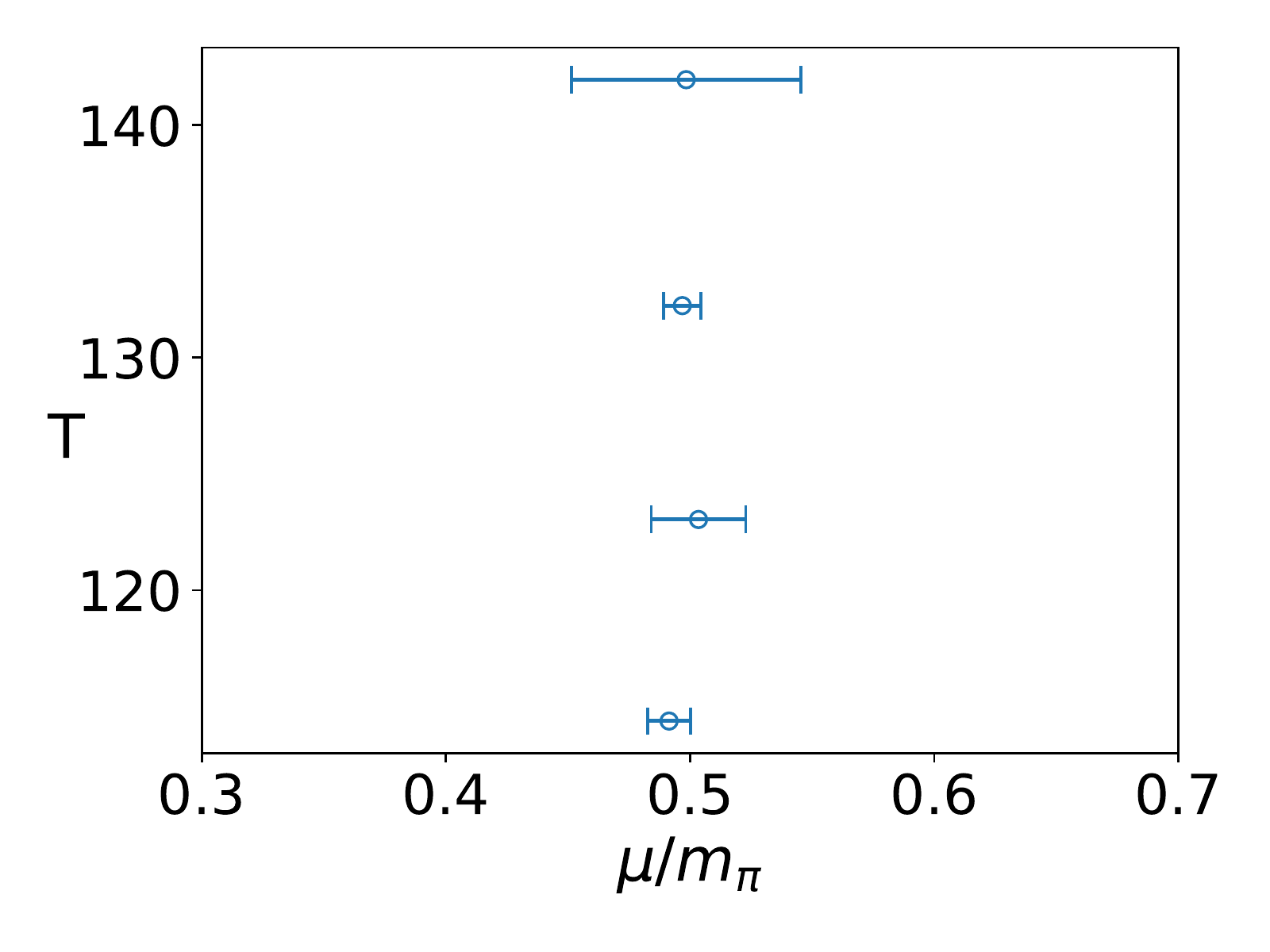} \\
\caption{The value of the improved pion condensate (left) and the 
 location of the pion condensation point (right) for four different values of
 temperature, obtained from the $24^3 \times 8$ lattice (no continuum extrapolation).}
\label{fig:pion-condensation-line}
\end{figure}

In Figure~\ref{fig:pion-condensation-line} we show our preliminary results 
on the value of the pion condensate and the location of the pion condensation line at four different values of the temperature. 
The interpolation was done using a cubic polynomial for the temperatures where we have enough points ($T=114$ MeV, $132$ MeV), and a quadratic polynomial otherwise.
Location of the transition point was determined as a point where the interpolation line intersects zero. 
In all cases all points giving a positive value of the pion condensate
were included in the fit. 
The error estimates for the fit line and the transition point were obtained 
by taking the standard deviation of the fit result for 100 simulated 
sets of condensate values normally distributed around the ``true'' values
with the known standard deviation. 

The results show that the vertical direction of the pion condensation 
line is preserved when going to smaller light quark masses, preferring the scenario shown in the left panel of the Figure~\ref{fig:chiral-pictures}: the location of the condensation line is compatible with the zero temperature location $\mu = m_\pi / 2$. 

As mentioned earlier, these results are based on the linear extrapolation to 
$\lambda=0$ from the data measured at nonzero values of the pion source parameter. 
To be able to check the validity of this extrapolation, we 
are currently performing further simulations at smaller values of $\lambda$.
Additionally, a $T$ scan at several values of $\mu_I > m_\pi / 2$ 
is being performed in order to locate the ``horizontal'' part of the pion condensation line as well.

\acknowledgments
This work was supported by the Deutsche Forschungsgemeinschaft (DFG, German Research Foundation) – project number 315477589 – TRR 211. The authors acknowledge the use of the Goethe-HLR cluster and thank the computing staff for their support.

\end{document}